\documentclass[10pt,conference]{IEEEtran}

\usepackage{graphicx}
\usepackage{cite}
\usepackage{url}
\usepackage{booktabs}
\usepackage{microtype}
\usepackage{subcaption}
\usepackage{amsmath}
\usepackage{boxedminipage}
\usepackage{balance}
\usepackage{color,soul}
\usepackage{listings}
\usepackage{setspace}

\newcommand{\tool}{\textit{Gradeer}}
\newcommand{\toolpossessive}{\textit{Gradeer's}}

\title{Gradeer: An Open-Source Modular Hybrid Grader}

\author{
	\IEEEauthorblockN{ 
		Benjamin S. Clegg\IEEEauthorrefmark{1}, 
		Maria-Cruz Villa-Uriol\IEEEauthorrefmark{1}, 
		Phil McMinn\IEEEauthorrefmark{1} and 
		Gordon Fraser\IEEEauthorrefmark{2}
	}
	\IEEEauthorblockA{\IEEEauthorrefmark{1}University of Sheffield, \IEEEauthorrefmark{2}University of Passau}
}

\begin{document}
	\maketitle
\begin{abstract}
	Automated assessment has been shown to greatly simplify the process of assessing students' programs.
	However, manual assessment still offers benefits to both students and tutors.
	We introduce \tool, a hybrid assessment tool, which allows tutors to leverage the advantages of both automated and manual assessment. 
	The tool features a modular design, allowing new grading functionality to be added.
	\tool~directly assists manual grading, by automatically loading code inspectors, running students' programs, and allowing grading to be stopped and resumed in place at a later time.
	We used \tool~to assess an end of year assignment for an introductory Java programming course, and found that its hybrid approach offers several benefits.
\end{abstract}
\section{Introduction}
The demand for Computer Science and Software Engineering education has continued to increase over recent years, with educational institutions seeing larger cohorts of students enrolled in such courses\cite{BCSPressOffice-IncreaseStudents}.
As technology further advances, future generations of students will drive this demand further, with universities and schools facing several challenges in teaching a growing number of students. 
One of these challenges is the assessment of a large number of students' solutions to programming tasks.
Assessment is particularly important, since it both has the ability to further students' development through the provision of detailed feedback, and serves to measure a student's understanding of a topic.

Automated grading and feedback techniques offer several benefits in assessing large numbers of students.
Their automated nature allows users to perform other tasks while grading is executed.
It is also often much quicker to run a series of automated processes than to manually assess individual students' solution programs. 
This is especially important for courses with large numbers of students, where manual assessment would consume too much time, and manual feedback could be provided too late to be of relevance to students' learning.
In addition, automated feedback allows for a large amount of feedback to be generated, and providing more pieces of automated feedback has been shown to improve students' performance~\cite{Falkner2014}.
Automated grading is also more consistent than manual grading, especially if students' solutions are assessed manually by multiple people~\cite{Albluwi2018}, which would likely be necessary to improve assessment times.

There are, however, some issues with the use of automated assessment alone.
There is a significant initial time cost of using automated assessment, with the need to either develop or configure a tool before assessment can be performed.
Additionally, with the exception of test-based systems, tutors may find it difficult to adapt an automated assessment system to meet their requirements~\cite{Keuning2016}.
Similarly, there are a wide range of unique automated assessment approaches~\cite{Liu2019, Insa2018, Singh2013, Parihar2017, Wunsche2018, Sridhara2016}, some of which may be suited to certain tasks, but would require a significant degree of effort to combine into one grading tool.
Automated assessment also lacks some of the benefits of manual approaches.
Manual assessment has the ability to capture aspects of grading that are hard to automate, such as the usefulness of variable names, or the appearance of a GUI. 
There is also evidence that manually provided feedback is of greater benefit to students' performance than automatically generated feedback~\cite{Leite2020}.

In this paper, we introduce \tool, a hybrid modular grading system, with the goal of providing the benefits of both approaches, while mitigating their challenges (Section~\ref{sect:gradeer}).
We used \tool~to assess an end of year assignment for an introductory programming course (Section~\ref{sect:case-study}).
We found that the tool's hybrid approach allowed for the use of a large number of consistent automated assessment criteria, and aided in the provision of detailed manual feedback to students.
\tool~also provides a degree of automation to assist tutors in manual assessment, such as automatically launching students' programs and code inspectors. 
We found that these features saved us a considerable amount of time when manually assessing students' solutions.
The modular nature of grading components allows a variety of automated grading techniques to be used in conjunction with one another, while minimising the effort required to combine their results.
\tool~is available on GitHub under the GPLv3 license, which allows users to write their own extensions and integrations for the tool\cite{GradeerRepository}.

	\section{The \tool~Grading Tool}
\label{sect:gradeer}

\tool~is an assessment tool which provides tutors with the benefits of both automated and manual assessment in a single package.
The tool achieves this using a modular design, allowing a user to choose how to assess a programming task using simple configuration files, or even define their own modules for specific purposes.
To allow for manual assessment, \tool~is designed to be used by tutors on personal computers, where the user can interact with the program via a CLI.
It is however possible for \tool~to be integrated with a GUI or web interface.
\tool~is implemented in Java, and allows for the assessment of Java programs. 
Wider language support is planned for future versions of the tool. 
This section describes our design of \tool, alongside some of its benefits.

\subsection{Checks}
\label{sect:gradeer-checks}

We designed \tool~with a focus on modular grading components, called \textit{checks}, each of which represents a single grading criterion.
Different types of checks are currently implemented, defining how a criterion's base score (a decimal value between zero and one) can be determined for a given process and student's solution.
Various checks of different types can be used together in a single run of \tool, constructing a markscheme to assess several learning outcomes.
For example, users can configure \tool~to use multiple checks to run various test suites, perform static analysis, and manually assess several aspects of a solution.
Users configure their checks in JSON files.
Users can also implement new checks to add the functionality of unique and domain-specific grading tools.

One currently implemented type of check is the \texttt{TestSuiteCheck}, which executes a given JUnit test class on a student's solution via Apache Ant~\cite{TheApacheSoftwareFoundation}, then calculates a score as the proportion of tests that pass. 
Tutors can assess individual learning outcomes by grouping tests that evaluate the same outcome into one class.

We also implemented check types for two static analysis tools, Checkstyle and PMD\cite{Checkstyle, PMD}, in order to automatically assess the code quality of students' solutions.
Such checks search the output of their respective tool for a user defined rule violation. 
The number of violations in each source file of a solution is recorded and used to compute a base score.
Users can also define a minimum and maximum number of violations, which yield base scores of one and zero, respectively.

To support manual assessment, we have implemented a \texttt{ManualCheck} type, which displays a user-defined prompt and score limit to the user when executed. 
This check then parses numeric input from the user and normalises it to a score in the range of zero and one.
For example, the following response would produce a base score of 0.6:
\begin{lstlisting}[linewidth=\columnwidth, breaklines=true]
How informative are the variable names? 
(0 = very poor, 10 = excellent)
# 6
\end{lstlisting}

Each check has an associated weight; a score multiplication factor to allow a test to have a greater or smaller impact on each solution's overall grade, as discussed in Section~\ref{sect:gradeer-output}.
This weight can be defined by the user. 
In addition, each check has associated feedback to provide to a student for their solution.
For most checks, this feedback is determined by mapping a base score to one of several feedback values that have been pre-defined by the user.
For example, the above manual check may provide students with feedback for the base scores, \textit{bs}:
\begin{itemize}
	\item $0.9\leq$~\textit{bs} $\leq1.0$: ``Your variable names are informative.''
	\item $0.5\leq$~\textit{bs} $<0.9$: ``Some of your variable names could be more informative.''
	\item $0.0\leq$~\textit{bs} $<0.5$: ``Most of your variable names could be more informative.''
\end{itemize}
Manual checks can also read text input from the user, allowing for additional feedback to be provided on an individual basis.
For example, a tutor may enter ``\texttt{a} is not an informative variable name, \texttt{leftMotor} would be better.''

\subsection{Execution}
\label{sect:gradeer-execution}
\begin{figure*}
	\centering
	\includegraphics[width=\textwidth]{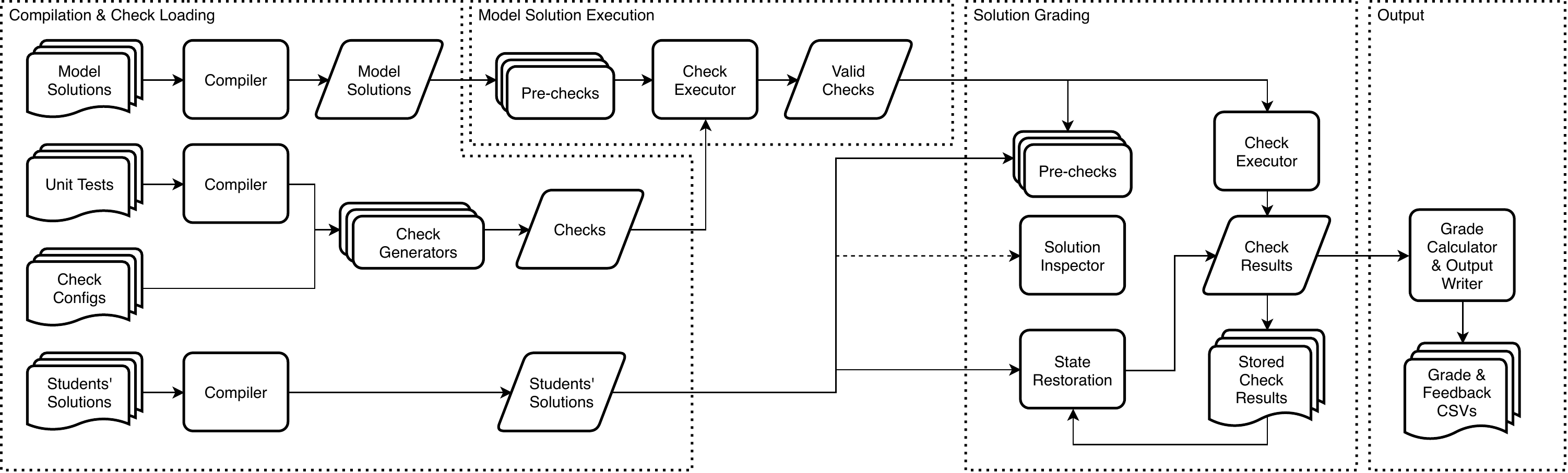}
	\caption{Overview of \toolpossessive~flow of execution. The dotted areas indicate different phases of the execution. Waved boxes are files, parallelograms are internal data, and regular boxes are processes.}
	\label{fig:gradeer-overview}
\end{figure*}

Figure~\ref{fig:gradeer-overview} shows an overview of \toolpossessive~execution process.

\subsubsection{Compilation \& Check Loading}
First, \tool~compiles every students' solution and every model solution (Section \ref{sect:gradeer-execution:model-solution}). 
At this stage, any solutions which do not compile are flagged as such. 
These solutions are reported to the tutor for review, and are excluded from further execution.
Next, \tool~loads every check defined in the JSON files. 
The tool also compiles the test classes that are provided by the user. 
If enabled, \tool~automatically generates a test suite check for each test class which does not have a matching check already defined by the user.

\subsubsection{Model Solution Execution}
\label{sect:gradeer-execution:model-solution}
The user can supply a set of one or more model solutions; entirely correct solutions to the programming task being assessed.
Users can choose to use multiple model solutions to define different correct implementations of the programming task.
In order to identify and remove invalid checks, \tool~executes every check on each provided model solution.
Checks which attain a base score of less than one on any of the model solutions are considered to be invalid, and are removed; they falsely claim that a model solution is partly or completely incorrect.
This prevents invalid checks from being used in the assessment of students' solutions, preventing them from unfairly losing or gaining grades, or being given inaccurate feedback.
For example, uncompilable test classes will not pass on any solutions, so their checks are removed.
The names of invalid checks are stored in a file for the tutor to review and correct.

\subsubsection{Solution Grading (for each Student's Solution)}

\paragraph{Pre-checks}
In order for some checks to function properly, a series of pre-checks are executed on each solution.
For example, checks for Checkstyle and PMD require pre-checks which execute their corresponding static analysis tool on the solution under test and store its output in memory. 

\paragraph{Solution Inspection}
To support effective manual grading, \tool~includes a \textit{solution inspector} which can perform two processes, as configured by the user.
The first executes a student's solution in a separate thread before running any manual checks. 
This allows the user to be able to interact with the solution, and to observe its user interface, which may be relevant to the rubric of manual checks.
The solution execution thread is closed following the completion of every manual check on a given solution.
The second opens each of the solution's source files in an external user defined text editor, such as Atom. 
This allows for the user to perform manual code inspection, for example to determine the quality of variable names or comments.
The solution inspector removes the need for the user to manually run a student's solution to interact with it, or open its source files to inspect it, saving time.

\paragraph{Checks}
The final step of a solution's grading process is to run every check on it. 
In order to improve execution time, \tool~runs automated checks in parallel by default. 
Manual checks are only executed in the main thread, however, as they require user input, and henceforth could result in the occurrence of race conditions otherwise.
In order to prevent some JUnit checks from taking too long to execute, \tool~has a user configurable global test timeout, where any tests that take longer than this time are treated as failing. 
This is particularly important, since some students' solutions may contain bugs that prevent them from halting, such as incorrect loop conditions.

\subsubsection{Output}
\label{sect:gradeer-output}

After executing every check on every solution, \tool~stores the appropriate grades and feedback for each solution in various CSV files.
The final grade of each solution is stored in one CSV file.
This grade is calculated by:
\begin{align*}
\text{\textit{Grade} (s)} &= \frac{\sum_{c\in C} w(c)\cdot\text{\textit{Base Score}} (c, s)}{\sum_{c\in C}w(c)}\text{,}\\
\text{where }s &= \text{Student's solution,}\\
C &= \text{Set of enabled checks,}\\
w(c) &= \text{Weight of check }c	
\end{align*}

Similarly, the combined feedback of each solution across all checks is also stored in a CSV file.
\tool~also stores a CSV file with the individual base scores and feedback of every check for each solution. 
This file also includes the weight of each check.
This allows for final changes to be made in spreadsheet software if absolutely necessary. 
For example, the user can post-process the students' grades by adjusting the checks' weights, and recalculating the final grades in the same manner as \tool.
Users can also gather valuable information on students' performance for the grading criteria, facilitating the provision of group feedback to the entire student cohort.

\subsection{State Restoration}
\label{sect:gradeer-state-restoration}
Following the completion of checks on a solution, \tool~stores the results and feedback for every check in a JSON file.
When \tool~is executed with such files present, it uses them to restore these check results for every applicable solution, and skips the corresponding checks when processing these solutions again. 
This has numerous advantages:

\begin{itemize}
	\item A tutor can effectively pause the grading process and come back to it at a later time. 
	This is particularly advantageous when using manual checks, as programming tasks with many students' solutions can take hours to manually assess.
	State restoration allows this arduous process to be split into more manageable marking sessions.
	\item Assessment tasks can be allocated to multiple users, such as TAs.
	Tutors can adjust users' \tool~configurations to use different solutions or checks. 
	By allocating different manual checks to different users, grading can be completed more quickly without reducing consistency. 
	By merging the users' JSON files and re-running \tool, the final grades and feedback can be generated.
	\item If \tool~halts unexpectedly, perhaps due to a wider system error, the user's grading progress is not lost.
	\item Tutors can either directly modify the result files to adjust the results of individual checks, or delete them outright to re-assess a solution. 
	Running \tool~again will update the final output files (as described in Section~\ref{sect:gradeer-output}).
	Tutors can also choose to add new checks after initial executions of the tool to capture additional assessment requirements. 
\end{itemize}


\section{Case Study}
\label{sect:case-study}
In this section, we discuss our application of \tool~in an end of year introductory Java programming assignment with 171 students' solutions.

\subsection{The Assignment}
The assignment required students to parse a series of structured input files into a provided data structure, then implement a set of methods that query this data. 
The assignment also required students to plot graphs using this data in a GUI using Java's Swing library.
A primary goal of the assignment was to provide students with experience in working on a multi-faceted project with codependent systems, which are more akin to real software than the simpler introductory programs used earlier in the course.
As an end of year assessment, the assignment had a fairly wide span of learning outcomes.
Such learning outcomes included the use of polymorphism, bespoke data structures, the choice and use of various Java Collections, text manipulation, GUI programming, algorithm design, and the use of good quality code and programming style.

We first determined the overall assignment specification, then focused on creating a model solution that captured this specification.
We then created a set of grading unit tests, ensuring that they were valid and that the model solution passed on each of them.
Following this, we duplicated the model solution to create a skeleton project, from which we removed the classes and methods that students were to implement.

\subsection{Release}

We distributed the skeleton project to students. 
We also provided the students with a set of input data files that were to be read by their implemented parsers. These data files were a subset of those that we later used when grading the assignment.
Around a week after we released the assignment, we also provided students with a set of public tests.
We designed these tests to ensure that students' code included the basic functionality of the assignment. 
This provided students with a degree of feedback as they worked on the assignment, and dissuaded students from submitting solutions which are not compatible with our grading environment, such as including incorrect class names.

\subsection{Check Configuration}

We configured \tool~to use 45 checks:
\begin{itemize}
	\item 26 test suite checks (each check executed one unit test),
	\item six PMD checks,
	\item six Checkstyle checks, and 
	\item seven manual checks (for GUI functionality and subjective aspects of code review, such as variable names).
\end{itemize}
By using these checks together, we were able to use \tool~to assess all of our learning outcomes.
The manual checks were important in this regard, since the design of the GUI and some aspects of code quality cannot be fully graded automatically. 

\subsection{Assessment}

While \tool~supports the use of all types of checks in a single execution, we split the checks across two separate execution configurations; one for automated checks and one for manual checks.
This was necessary since we anticipated that some solutions would be problematic, containing issues that would prevent compilation or execution.
As such, running manual checks on some of these solutions would have been a waste of effort if the solutions could not be executed properly.
By splitting the checks we were able to first compile the students' solutions and run the automated checks to identify any problematic solutions, and to assess the working solutions.
We identified 48 problematic solutions. 
We repaired these solutions where possible so that they could still be graded with \tool, but added a penalty for doing so when post-processing the grades.
We repeated the automated grading for these repaired solutions.
However, 11 of the solutions could not be repaired due to severe issues.
We wrote individual feedback for each of these solutions to explain the nature of these problems.
Finally, we re-executed \tool~with only the manual checks on every working and repaired solution.
Table~\ref{tbl:times} shows the average amount of time that various aspects of running the assessment with \tool~took for each applicable solution, alongside the time taken to manage problematic solutions.
\begin{table}[]
	\centering
	\caption{Average time to perform each assessment task on each applicable solution.}
	\label{tbl:times}
	\begin{tabular}{@{}ll@{}}
		\toprule
		\textbf{Assessment Task}            & \textbf{Average Time Per Solution} \\ \midrule
		\textit{Compilable Solutions}       &                                    \\
		Compilation                         & $\sim$1.6 seconds                  \\
		38 Automated Checks                 & $\sim$28.2 seconds                 \\
		7 Manual Checks                     & $\sim$2 minutes                    \\ \midrule
		\textit{Problematic Solutions}      &                                    \\
		Problem Identification              & $\sim$11.3 minutes                 \\
		Solution Repair     & $\sim$11.4 minutes                 \\
		Individual Feedback & $\sim$10 minutes                   \\ \bottomrule
	\end{tabular}
\end{table}

Once we completed grading the assignments, we 
performed some post-processing on the results.
In particular, we added some more specific feedback and adjusted the weights of some of the checks.
Providing the additional feedback revealed the possible benefit of being able to add specific feedback when running \tool, leading us to later implement the ability to add user entered feedback for manual checks.
We also provided more detailed and general feedback to the entire student cohort using the distribution of solutions' base scores for individual checks.
In addition, we used this check performance data to adjust the checks' weights.
For example, we found that the scores of some checks would vary considerably between solutions, such as a PMD check for cyclomatic complexity, for which approximately half of the solutions achieved $<0.5$.
In such cases, we increased the check's weight, as it better differentiated students' solutions.
However, we attempted to maintain similar total weights between the broader groups of learning outcomes, such as overall correctness and code quality, to assess students in a well-rounded manner.

\subsection{Benefits of \tool}
We found that \toolpossessive~hybrid grading approach provided several benefits when assessing this programming assignment:

\subsubsection{Fast Manual Assessment}
\tool~provides a particular benefit in allowing for quick manual assessment.
This is mostly due to \toolpossessive~solution inspector, which automatically executes students' solutions, and displays their source files in a text editor.
Without this feature, a tutor must manually open the correct directory, enter a command to run the solution, and open the source files, before beginning the manual assessment.  
By removing the need to follow these steps for every solution, \tool~removes a significant bottleneck in manual grading.

\subsubsection{Automated Grading}
By using automated grading wherever possible, we were able to reduce the number of manual checks.
For example, we used some static analysis checks to evaluate the style of students' solution programs, such as ensuring that they used camel case formatting in variable names. 
By using these checks, the tutor did not have to look for these issues when performing the manual code inspection.
Similarly, the use of unit tests to assess correctness of some elements of the program removed the need for the tutor to identify faults in these elements manually. 
The additional benefit of automated grading is that the checks are applied consistently across solutions. 
Any two students' solutions which have the same faults will be assessed the exact same way.

\subsubsection{High Quality Feedback}
We found that \tool~was capable of providing useful feedback to students.
While automated checks only provide simple feedback, the large number of these checks gave students a very wide range of feedback; they could gain a good understanding of where they succeeded and where they can improve.
This is supported by Falkner et al.'s findings that students' performance improves as more pieces of automated feedback are provided~\cite{Falkner2014}.
This feedback is further augmented by \toolpossessive~support for manual checks, the scores of which we used to determine which of several pieces of feedback to give to a student. 
The ability to provide manual feedback at runtime in the current version of \tool~supports this even further.

\subsubsection{Reusable}
In the past, we typically used unique autograding scripts for each assessment. 
Developing these scripts is a time consuming process, and may involve repeated effort of implementing similar functionality across multiple assessments.
Conversely, \tool~can be reused in different assessments, only requiring modifications to simple configuration files.

\subsection{Challenges}
When assessing the assignment, we found that uncompilable solutions introduced the greatest time cost. 
Around 48 of the 171 solutions initially could not be compiled or executed, due to missing files, syntax errors, or modifying files that should be unmodified.
It is possible that such problems could be mitigated by preventing students from uploading broken solutions, such as by integrating \tool~with the solution upload system, and reporting to students if an issue is detected.

Running the automated checks did take a considerable amount of time, at $\sim$28.2 seconds per solution using an AMD Ryzen 1700 CPU.
The main source of this time cost is setting up the test execution environment. 
We plan to investigate a possible workaround for this issue in the future.
In addition, the version of \tool~that we used for this assessment did not support multithreading.
After implementing multithreading, we observed an execution time of $\sim$10.9 seconds per solution.

We found that some static analysis rules can present a unique challenge in being used in an automated grader.
In particular, Checkstyle's indentation rules can only be used with one tutor defined indentation width, while indentation widths may vary between solutions. 
This is an issue since several different indentation widths are commonly used in software, any of which may be acceptable provided that they are used consistently. 
It may be possible to use multiple similar checks and only use the highest base score as a workaround.

While using software such as \tool~requires less effort than writing a unique grading script, some tutors may be dissuaded by not understanding its internal functionality. 
Providing tests may increase tutors' confidence in such tools.

\section{Related Work}

Some existing automated grading tools also feature modular assessment elements~\cite{Zschaler2018}.
For example, Nexus's assessment components implemented as Docker micro-services~\cite{Zschaler2017}.
Web-CAT uses modular plug-ins~\cite{Edwards2008, Edwards-WhatIsWebCAT}.
JACK and ArTEMiS both use multiple software components that can be split across multiple servers, and interchanged to support different grading functionalities~\cite{Goedicke-2008, Krusche2018b}.
These tools are designed to be used as scalable web services, which can be beneficial for large courses and MOOCs. 
Such approaches do have considerable advantages, and may allow tutors to view students' source code, but tutors cannot run and interact with students' solutions directly, which limits their ability to perform manual assessment.
By contrast, \tool~specifically accommodates manual assessment.

It is not uncommon for assessment tools to take a ``semi-automatic'' approach, with support for user intervention and manual assessments alongside automated processes~\cite{Souza2016}. 
Web-CAT allows tutors to manually inspect students' source code, and provide feedback or additional grades~\cite{Edwards-WhatIsWebCAT}.
Praktomat grants TAs the ability to provide manual feedback by adding comments to students' code~\cite{Brietner-2017}. 
It also allows TAs to add manual scores for learning outcomes.
JACK enables tutors to provide manual corrections for generated grades, and manual feedback~\cite{Goedicke-2008}.
Jackson's grading tool displays the contents of a solution's files before reading the user's input to determine the scores of a series of manual assessment elements~\cite{Jackson2000}. 
While these tools have provisions for manual assessment, none of them automate the process of launching students' programs for tutors to interact with them.
This may be problematic, as the bottleneck of manually running each solution is still present when evaluating user interaction.
\toolpossessive~solution inspector removes this bottleneck entirely.
\tool~also combines the results of automated and manual checks into a single grade, without additional user intervention.
\section{Conclusions and Future Work}

In this paper we have presented \tool, a modular grading tool to support both the automated and manual assessment of students' programs. 
We have also discussed our experiences in using the tool to assess an end of year assignment for an introductory programming course.
We find that \tool~can effectively support tutors in providing quality feedback to students, while maintaining a low time cost of assessment.
\tool~also provides tutors with detailed data on students' performance, which can be used to inform and improve teaching quality, future assessment design, and feedback.
\tool~is available at \url{https://github.com/ben-clegg/gradeer}~\cite{GradeerRepository}.

In our future work, we will extend our evaluation of \tool, by comparing the time saved using our solution inspector versus manually running each solution, and by surveying more end users.
We plan to improve \tool, such as enhancing its modularity, by further separating check modules from the rest of the system, and modularising other components (such as pre-checks and language-specific functionality) as well.
We also intend to add web integration to the tool, to inform students when they have submitted solutions with significant problems.
	\newpage
	
	
	\balance
	\bibliography{gradeer-paper}
	\bibliographystyle{ieeetr}
\end{document}